\documentclass[a4paper]{article}
\usepackage{INTERSPEECH2020}
\usepackage{tikz,tikz-qtree}
\usepackage{amsmath,amssymb,multirow,url}
\usepackage[utf8]{inputenc}

\title{Context Dependent RNNLM for Automatic Transcription of Conversations}
\name{Srikanth Raj Chetupalli, Sriram Ganapathy\thanks{This work was funded by British Telecom India Research Center (BTIRC) project on Speech Analytics.}}
\address{ LEAP Lab, Indian Institute of Science, Bengaluru, India}
\email{sraj@iisc.ac.in, sriramg@iisc.ac.in}

\begin{document}
\maketitle
\begin{abstract}
Conversational speech, while being unstructured at an utterance level, typically has a macro topic which provides larger context spanning multiple utterances. The current language models in speech recognition systems using recurrent neural networks (RNNLM) rely mainly on the local context and exclude the larger context. In order to model the long term dependencies of words across multiple sentences, we propose a novel architecture where the words from prior utterances are converted to an embedding. The relevance of these embeddings for the prediction of next word in the current sentence is found using a gating network. The relevance weighted context embedding vector is combined in the language model to improve the next word prediction, and the entire model including the context embedding and the relevance weighting layers is jointly learned for a conversational language modeling task. Experiments are performed on two conversational datasets - AMI corpus and the Switchboard corpus. In these tasks, we illustrate that the proposed approach yields significant improvements in language model perplexity over the RNNLM baseline. In addition, the use of proposed conversational LM for ASR rescoring results in absolute WER reduction of $1.2$\% on Switchboard dataset and $1.0$\% on AMI dataset over the RNNLM based ASR baseline. 

\end{abstract}

\section{Introduction}
Language modeling (LM) is the task of predicting the next word given the past history of words in a text stream and it forms an integral part of automatic speech recognition (ASR) systems and other natural language processing systems. In the past decade, following a similar trend in several other domains, the methods used for LM have shifted fundamentally from $n$-gram models based on frequency of counts to deep neural network based models. The earliest approach to LM using feed-forward networks was proposed by Bengio et al.~\cite{Bengio2003Neural}. These models were advanced using  recurrent neural networks (RNNs) by Mikolov et. al~\cite{mikolov2010recurrent} and then further using long short-term memory (LSTM) variants of RNNs by Sundermeyer et al.~\cite{sundermeyer2012lstm} and more recently by Xiong et al.~\cite{xiong2017toward}. The RNN models have multiple advantages over the traditional $n$-gram framework. The continuous-space embedding of words allows word similarities to be computed in an efficient manner for generalization~\cite{mikolov2010recurrent}, and the recurrent architecture also allows an unlimited history to condition the prediction of next word. 
\par In the current implementation of LM (even for conversations), the potential advantage of unlimited history, however, is not used to its full extent. The LM is typically ``reset'' at the start of each utterance in current state-of-the-art ASR systems \cite{saon2017english, xiong2017toward}. This approach assumes that the successive utterances are independent of each other, which is not the case in conversational data. In the past, there have been some attempts to incorporate the information from a larger context, spanning multiple speaker utterances, in $n$-gram models (for example, Bellegarda et. al~\cite{bellegarda2004statistical} and Ji et.al~\cite{ji2004multi}). For neural network based LM, the inclusion of a longer context as a slowly varying context vector was attempted by Mikolov et.al.~\cite{Mikolov2012Context} where the context vector was incorporated as a latent semantic embedding of the previous context. Xiong et.al~\cite{xiong2018session} recently proposed a LSTM based LM that takes the history of the conversational utterances in the LSTM model. However, the modeling of long term dependencies spanning multiple sentences can pose a substantial challenge in LSTM models. In addition, all the words of the current sentence may not benefit from the inclusion of the longer context.

\par In this paper, we propose a novel approach to LM, referred to as context dependent RNNLM (CRNNLM), using the word embeddings from contextual sentences along with the words of the current sentence. The cross correlation of the word embeddings from the contextual sentences with the embedding of the current word is used to derive a context embedding. This contextual embedding is used in a gating network to generate a relevance weighted context vector that is combined with the word embeddings of the current sentence in the LM. Unlike the approach in \cite{Mikolov2012Context}, where a single context vector is computed for the sentence, a separate context vector is computed for each word in the current sentence to capture local context. The entire model, including the gating network and embedding layers, is learned jointly from the training data corresponding to conversational speech. Experiments on Switchboard dataset~\cite{switchboard} and AMI meeting dataset~\cite{AMIMeetings} show that the proposed approach to LM improves significantly over the state-of-art RNNLM in terms of perplexity measure, and also in terms of WER in the ASR task. 
\par The rest of the paper is organized as follows. Section~\ref{sec:crnnlm} describes the proposed model to LM using contextual embeddings. The experiments and results using the proposed LM are reported in Section~\ref{sec:expt},
which is followed by a brief summary of the work in Section~\ref{sec:concl}.


\section{Context dependent RNNLM}\label{sec:crnnlm} 
\begin{figure}[t]
    \centering
    \begin{tikzpicture}[scale=0.7, every node/.style = {scale=0.7}]
        \node[draw=red,rounded corners,text width=0.7in,align=center] at (0,0) {speaker-1};
        \node[draw=blue,rounded corners,text width=0.5in,align=center] at (1.2,-0.8) {speaker-2};
        \node[draw=red,rounded corners,text width=0.5in,align=center] at (2.8,0) {speaker-1};
        \node[draw=blue,rounded corners,text width=0.8in,align=center] at (4.5,-0.8) {speaker-2};
        \node[draw=red,rounded corners,text width=0.5in,align=center] at (6.2,0) {speaker-1};
        \draw[->] (-1.2,-1.2) --node[above,pos=0.95]{time} (7.5,-1.2);
        \draw[-] (-1.2,-1.2) -- (-1.2,0.5);
        
        \draw[->,double] (3.5,-1.3) -- (3.5,-1.9);
        
        \node[draw=red,rounded corners,text width=0.7in,align=center] at (0,-2.1) {speaker-1};
        \node[draw=blue,rounded corners,text width=0.5in,align=center] at (1.8,-2.1) {speaker-2};
        \node[draw=red,rounded corners,text width=0.5in,align=center] at (3.35,-2.1) {speaker-1};        
        \node[draw=blue,rounded corners,text width=0.8in,align=center] at (5.3,-2.1) {speaker-2};
        \node[draw=red,rounded corners,text width=0.5in,align=center] at (7.2,-2.1) {speaker-1};
    \end{tikzpicture}
    \vspace{-5pt}
    \caption{Conversation progression and utterance serialization.}
    \label{fig:utterance_strct}
\end{figure}
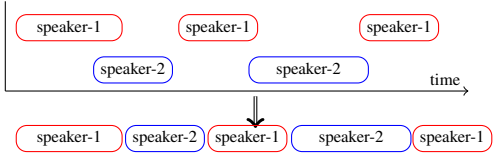

A block diagram of the proposed context dependent RNNLM (CRNNLM) is shown in Fig. \ref{fig:crnnlm_bd}. The utterances in the conversation are serialized, based on the onset time, as shown in Fig. \ref{fig:utterance_strct}. To model a given utterance, a set of past $c$ utterances is considered as the context. The current utterance and the context utterances are given as two separate input streams to the neural network. The neural network is divided into three stages. In the first stage, a hidden representation is computed for each word in the input and context streams. In the second stage, a cross-attention like operation \cite{vaswani2017attention} between input and context streams, is used to derive a context embedding separately for each word in the input stream. The context embedding is further gated using a relevance weight, which quantifies the importance of context in predicting the next word. In the final stage, the relevance weighted context embedding is combined with the input hidden representation and fed to a next word prediction network. A detailed description of the different computation steps is summarized below.
\begin{figure}[t]
    \centering
    \begin{tikzpicture}[scale=0.75, every node/.style = {scale=0.75}]
        \node[draw=blue,text width=1.2in,align=center,rounded corners] (we) at (2.3,0) {Word embedding (${\bf E}_i$)};
        
        \draw[->,blue,double] ([yshift=-15pt]we.south) --node[below,pos=0.25]{\large $<s>,w_1,\dots,w_{N-1}$} (we.south);
        
        \node[draw=red,text width=1.2in,align=center,rounded corners ] (ae) at (-2,0) {Word embedding (${\bf E}_i$)};
        
        \draw[->,red,double] ([yshift=-15pt]ae.south) --node[below,pos=0.25]{\large $a_1,\dots,a_{L}$} (ae.south);        
        \node[draw=blue,text width=1.2in,align=center,fill=blue!20,rounded corners] (ihr) at (2.3,1.0) {Hidden Representation};
        \node[draw=red,text width=1.2in,align=center,fill=red!20,rounded corners] (ahr) at (-2,1.0) {Hidden Representation};
        
        \draw[->,blue,double] (we.north) --node[right]{\large $\{{\bf w}_{t} \} \in \mathcal{R}^{1024 \times N}$} (ihr.south);
        \draw[->,red,double] (ae.north) --node[left]{\large $\{{\bf a}_{l} \}  \in \mathcal{R}^{1024 \times L}$} (ahr.south);
        
        \node[draw=red!50!blue,text width=1.2in,align=center,minimum height=0.2in,rounded corners] (cc) at (-2,2.5) {Cross relation embedding};
        
        \draw[->,red,double] (ahr.north) --node[left]{\large $\{{\bf g}_{l}\}  \in \mathcal{R}^{256 \times L}$} (cc.south);
        
        \draw[->,double,blue] (ihr.north) |- (cc.east); 
        
        \node[draw=red!50!blue,text width=1.1in,align=center,minimum height=0.2in, rounded corners] (sg) at (-2,3.8) {Relevance weighting};
        
        \draw[->,double,blue] (ihr.north) |- (sg.east); 
        
        \draw[->,double,red!50!blue] (cc) --node[left]{\large $\{ {\bf c}_{t}\}  \in \mathcal{R}^{256 \times N}$} (sg);
        
        \node[draw=cyan,text width=1.2in,align=center,minimum height=0.2in, rounded corners] (FA) at (2.3,5.2) {Feature combination};
        
        \draw[->,double,red!50!blue] (sg.north) |-node[left,pos=0.2]{\large $\{\hat{\bf c}_{t}\}  \in \mathcal{R}^{256 \times N}$} (FA.west);        
        \draw[->,double,blue] (ihr.north) --node[right,pos=0.1]{\large $\{ {\bf h}_{t} \}  \in \mathcal{R}^{256 \times N}$} (FA.south);
        
        \node[draw=cyan,text width=1.2in,align=center,minimum height=0.2in,fill=cyan!20, rounded corners] (nwp) at (2.3,6.3) {Next word predicion};
        
        \draw[->,double,cyan] (FA.north) --node[right]{\large $\{ \hat{\bf h}_{0:N-1} \}$} (nwp.south);
        
        \node[draw,text width=1.2in,align=center,minimum height=0.2in,rounded corners] (ope) at (2.3,7.4) {Output Embedding ${\bf E}_o$};     
        
        \node[draw,text width=1.2in,align=center,minimum height=0.2in,rounded corners] (sm) at (2.3,8.5) {Softmax};             
        \draw[->,double,cyan] (nwp.north) --node[right]{\large $\{ \tilde{\bf o}_{0:N-1} \}  \in \mathcal{R}^{1024 \times N}$} (ope.south);
        
        \draw[->,double,black] (ope.north) --node[right]{\large $ \{ {\bf o}_{0:N-1} \}  \in \mathcal{R}^{|V| \times N} $} (sm.south);
        
        \draw[->,double,black] (sm.north) --node[above,pos=0.7]{ \large $w_1,w_2,\dots,w_N$} ([yshift=15pt]sm.north);
        
        \node[draw=blue,text width=0.65in,fill=blue!20,rounded corners, align=center] at (4.25,3) {LSTM + linear + tanh};
        \node[draw=red,text width=0.65in,fill=red!20,rounded corners, align=center] at (-2,6.5) {BiLSTM + linear + tanh};
        \node[draw=cyan,text width=0.65in,fill=cyan!20,rounded corners, align=center] at (-2,8) {LSTM + linear};
        
        \draw[dashed,rounded corners] (0.4,-0.4) rectangle (4.8,1.4);
        \draw[dashed,rounded corners] (-4.8,-0.4) rectangle (0.1,1.4);
        \draw[dashed,rounded corners] (-4.8,1.9) rectangle (0.1,4.25);
    \end{tikzpicture}
    \vspace{-5pt}
    \caption{Block diagram of the C-RNNLM architecture.}
    \label{fig:crnnlm_bd}
\end{figure}
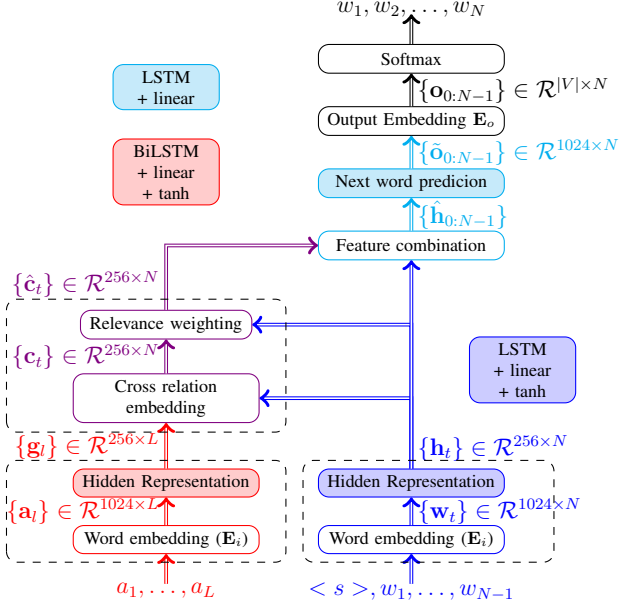

\subsection{Input representation}
Let $\{ {\bf w}_0,\dots,{\bf w}_{N-1} \}$ denote the sequence of word embeddings of the current input sentence containing words $<s>,w_1,\dots,w_{N-1}$. The hidden representation ${\bf h}_t$ for the $t^{th}$ input word is obtained via the following operations:
\begin{equation}
    {\bf h}_t = tanh\left( \mbox{linear} \left( \mbox{LSTM} \left( \{ {\bf w}_t \} \right) \right) \right),~\forall~t.
\end{equation}
At time instant $t$, the uni-directional LSTM computes a representation dependent on the words $w_{\leq t}$, which is linearly projected and the tanh activation is used for range compaction. The word embedding ${\bf w}_t$ is obtained from 1-hot-K representation of input word $w_t$ using the input embedding matrix ${\bf E}_i$. We use the transpose of the output embedding matrix ${\bf E}_o$ as the input embedding matrix, i.e., ${\bf E}_i = {\bf E}_o^T$, inspired by \cite{press2017using}. 
\subsection{Context embedding}
Let $L$ denote the number of words from context (previous $c$) utterances, and let $\{ \bf{a}_1,\dots,\bf{a}_L \}$ be the sequence of corresponding word embeddings computed using the input embedding matrix ${\bf E}_i$ shown in Fig. \ref{fig:crnnlm_bd}. The hidden representation ${\bf g}_l$ for the $l^{th}$ word in the context input is computed as:
\begin{eqnarray}
    {\bf g}_l = tanh\left( \mbox{linear}\left( \mbox{BiLSTM}\left(  \{{\bf a}_l\} \right) \right)\right),~\forall~l.
\end{eqnarray}
The BiLSTM layer looks at all the words in the context, and the linear layer with tanh activation is used for joint projection of the forward and backward LSTM output streams and range compaction. The context embedding $\bf{c}_t$ corresponding to the word $w_t$ is then computed as a weighted sum of representations $\{ {\bf g}_1,\dots,{\bf g}_L \}$ of context input.
\begin{equation}
    {\bf c}_t = \sum_{l=1}^{L} \alpha_{tl} {\bf g}_l,
\end{equation}
where the weight vector $\boldsymbol{\alpha}_t = \left[ \alpha_{t1},\dots,\alpha_{tL} \right]^T$ is computed using cross correlation as,
\begin{equation}\label{eqn:alphat}
    \boldsymbol{\alpha}_{t} = softmax( {\bf h}_t^T [{\bf g}_1,\dots,{\bf g}_L] ).
\end{equation}
The correlation between ${\bf h}_t$ and ${\bf g}_l$ will be high, if the corresponding words are related, or occur frequently in a similar context. Hence, the derived context embedding ${\bf c}_t$ captures local context information.
\subsection{Relevance weighted context embedding}
The context may not be relevant to predict the next word in a word sequence. To model this, we consider relevance weighting of the context embedding using joint, linear projection of the context embedding and the input representation. This method is inspired by the cold fusion method proposed in \cite{Sriram2018}. We explore two strategies for relevance weighting, (i) coarse weighting and (ii) fine weighting.
\par In coarse weighting, the context embedding is multiplied by a {\it scalar} weight: $\hat{\bf c}_t = {\beta}_t {\bf c}_t$, where
    \begin{equation}
        {\beta}_t = sigmoid\left( {\bf w}^T \left[ {\bf h}_t^T ; {\bf c}_t^T \right]^T  \right),~{\bf w} \in \mathcal{R}^{2h\times 1},
    \end{equation}
and in fine weighting, the context embedding is multiplied element-wise by a {\it vector} weight: $\hat{\bf c}_t = {\boldsymbol \beta}_t \odot {\bf c}_t$, where
    \begin{equation}
        {\boldsymbol \beta}_t = sigmoid\left( {\bf W}^T \left[ {\bf h}_t^T ; {\bf c}_t^T \right]^T  \right),~~{\bf W} \in \mathcal{R}^{2h\times h}.
    \end{equation}
\subsection{Feature combination}
    We explore additive and concatenative schemes for combining the relevance weighted context embedding vector with the input feature. In the additive scheme, the combined feature is obtained as,
    \begin{equation}
        \hat{\bf h}_t = {\bf h}_t + \hat{\bf c}_t,
    \end{equation}
    and in the concatenative scheme, the combined feature is obtained as,
    \begin{equation}
        \hat{\bf h}_t = \left[  {\bf h}_t^T ; \hat{\bf c}_t^T \right]^T.
    \end{equation}
\subsection{Next word prediction and output embedding}\label{sec:nwp}
The next word prediction network consists of a uni-directional LSTM layer followed by a linear layer, which gives an embedding vector as the output $\tilde{\bf o}_t$:
\begin{equation}
    \tilde{\bf o}_t = \mbox{Linear} \left( \mbox{LSTM} \left( \hat{\bf h}_t \right) \right)
\end{equation}
The output embedding matrix ${\bf E}_o$ is used to project the embedding $\tilde{\bf o}_t$ to the word level output ${\bf o}_t$ 
$\left( {\bf o}_t = {\bf E}_o \tilde{\bf o}_t \right)$.
The softmax of the vector ${\bf o}_t$ is taken as the word level posterior distribution for the next word prediction.
\section{Experiments and results}\label{sec:expt}
We consider evaluation of the proposed language model using two datasets, (i) Switchboard (SWB) telephone conversation dataset ($300$ hour subset) \cite{switchboard}, and (ii) AMI meeting dataset \cite{AMIMeetings}. The CRNNLM is trained separately for each dataset, and uses the same architecture. The vocabulary size is $30278$ words for the Switchboard dataset and $46950$ for the AMI dataset. The number of training conversations is $2405$ and $130$, for the Switchboard and AMI datasets respectively. Further, we augment the training set with $11699$ Fisher conversation transcripts \cite{FisherTranscripts} for both the setups. The total number of sentences and words in the training set is ($3$M,$31.7$M),~($2.5$M,$24.8$M) for the Switchboard and AMI setups respectively. The evaluation set consists of $40$ conversations in the Switchboard and $10$ conversations in the AMI dataset. The training set is divided into $50$ subsets with approximately equal number of conversations. The LM is trained with cross entropy loss using Adam optimizer \cite{kingma2014adam}. The initial learning rate is chosen to be $0.01$, and decreased with a constant decay coefficient of $0.99$ after every subset (starting from $20^{th}$ subset). The batch size is chosen to be $32$ samples. The length of input training examples is restricted to a maximum of $16$ words, and the context example length is not constrained. A $\ell_2$ regularization with a weight of $10^{-5}$ is applied to all the weights in the network, and dropout regularization with $p=0.2$ is used for the LSTM layer outputs. The CRNNLM is trained using the PyTorch toolkit \cite{NEURIPS2019_9015}. 
\par We explore four variants of the CRNNLM architecture, which are summarized in Table~\ref{tab:crnnlmarchs}. We also compare CRNNLM with the trigram LM, RNNLM implemented using standard Kaldi (Kaldi-RNNLM) recipe, and the session level language model (SessionLM) proposed in \cite{xiong2018session}. For the SessionLM, we choose three hidden layers as in \cite{xiong2018session}, but with $256$ units each, and the embedding used to encode the word inputs is trained jointly with the LM. We consider the variant of SessionLM in \cite{xiong2018session} without the speaker change and speaker overlap information, and the network is trained similar to CRNNLM. During training words from previous $10$ sentences are used as session history and during evaluation all the past sentences in the conversation are used as session history.
\subsection{Discussion and analysis}
\begin{table}[t]
    \centering
    \caption{CRNNLM architecture variants}
    \vspace{-5pt}
    \begin{tabular}{|c|c|c|}
        \hline
        Name & Relevance weight & Feature combination \\\hline
        V1 & Unity & Concatenative\\
        V2 & Scalar & Concatenative\\
        V3 & Vector & Concatenative\\
        V4 & Vector & Additive\\\hline
    \end{tabular}
    \label{tab:crnnlmarchs}
    \vspace{-10pt}
\end{table}
\begin{figure}[t]
    \centering
    \includegraphics[width=2.7in,height=1.5in,trim={0 0 0 20},clip]{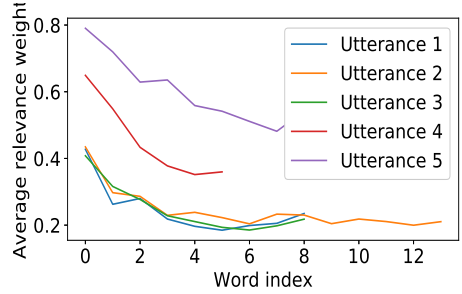}
    \vspace{-5pt}
    \caption{Relevance weight $\boldsymbol{\beta}_{t=1:N}$ averaged across the context embedding dimension, for the CRNNLM architecture V3, for five different utterances from Switchboard evaluation set.}
    \vspace{-5pt}
    \label{fig:rwutts}
\end{figure}
\par First, we investigate the importance of context embedding, using the CRNNLM architecture V3 trained on Switchboard dataset. Fig. \ref{fig:rwutts} shows the average relevance weight $\boldsymbol{\beta}_t$ for $5$ different test utterances. A higher value indicates the network is giving importance to the context embedding. We see that, on an average, a higher weight is given to the context embedding in the prediction of the first few words of the current utterance, and the context becomes less relevant gradually. Also, there is a difference across the utterances, indicating that the context embedding can be more relevant in certain local context scenarios. To explore this further, we study the perplexity of the prediction of first word in the utterance.
\par
\begin{figure}[t]
    \centering
        \includegraphics[width=2.6in,height=1.5in]{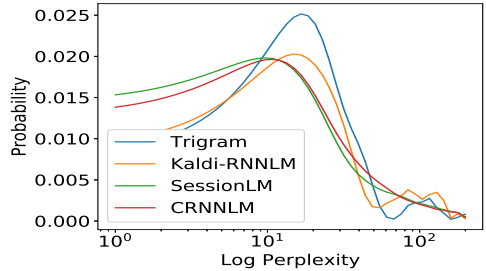}
    \vspace{-5pt}
    \caption{Comparison of the PDF (kernel density estimate) of first word perplexity, i.e., $1/P(word | <s>)$ for all utterances in the evaluation set of Switchboard dataset. 
    }
    \vspace{-5pt}
    \label{fig:wordprobability}
\end{figure}

 Figure \ref{fig:wordprobability} shows a comparison of the probability density function (PDF) of the first word perplexity (FWP). We see that, the PDF is shifted towards zero for CRNNLM (network V3) and SessionLM, indicating better first word prediction compared to trigram or Kaldi-RNNLM. The min value of FWP for trigram and Kaldi-RNNLM is found to be $8$ and $9.43$ respectively, but CRNNLM predicts words with a higher probability in certain contexts, and the min FWP is $1.08$.
\begin{figure}[t!]
    \centering
    \includegraphics[width=3.0in,height=1.5in,trim={0 50 0 110},clip]{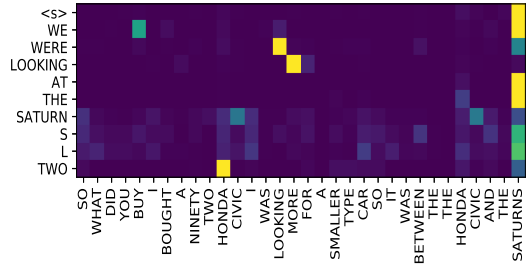}
    \vspace{-5pt}
    \caption{Cross relation weights $\boldsymbol{\alpha}_t$ (eqn. \eqref{eqn:alphat}) computed in the CRNNLM architecture V3. The x-axis corresponds to the words in context utterances from both the speakers.}
    \vspace{-5pt}
    \label{fig:crossAttentionWeights}
\end{figure}
\par Fig. \ref{fig:crossAttentionWeights} shows the cross relation weights $\boldsymbol{\alpha}_t$ of eqn. \eqref{eqn:alphat} for the sentence ``We were looking at the Saturn S L two'', which was part of a Switchboard conversation. For the prediction of first word (``We''), i.e., for $<s>$ as input, a relatively higher weight is given to the last word in the context input, indicating that the network is trying to ensure continuity across the sentences. At $t=3$, `$<s>$ We were' is given as input, a higher weight is given to the word `looking' in the context input which is also the correct next word; and in the next time step, the weight is high for the `more', which corresponds to the sub-sequence `looking more' in the context input. For the subsequent time steps, the weights are spread across all words with relatively higher weights for the words \{`Honda', `Civic',`Saturns',`Car'\}, indicating the computation of an average embedding with a preference to the macro topic of the conversation. Overall, we observe that, the CRNNLM approach promotes continuity across utterances, and gives higher importance to sub-sequences from context input to improve the next word prediction. 
\par We study the LM performance using perplexity measure. Table \ref{tab:lmperplexity} shows the LM perplexity on the evaluation set. We see that, neural LMs (CRNNLMs, Kaldi-RNNLM and SessionLM~\cite{xiong2018session}) show significant improvement in perplexity compared to the trigram LM. The performance of CRNNLM is better than the Kaldi-RNNLM and the SessionLM~\cite{xiong2018session}. Comparing architecture V1 with V2-V4, we see that relevance weighting improves the LM perplexity. The feature combination method (concatenative or additive), and the relevance weighting scheme (coarse or fine) are found to have a minor effect on perplexity.
\begin{table}[t]
    \centering
    \caption{LM Perplexity of various models.}
    \vspace{-5pt}
    \label{tab:lmperplexity}
    \begin{tabular}{|c|c|c|}
    \hline
    \multirow{2}{*}{Net} & \multicolumn{2}{c|}{Perplexity} \\ \cline{2-3} 
                         & SWB        & AMI        \\ \hline
                        trigram & 112.32  & 124.05     \\\hline
                        Kaldi-RNNLM & 93.00  & 100.22    \\\hline
                        SessionLM~\cite{xiong2018session} & 61.94  & 79.28    \\\hline V1 &    63.60   &   72.99       \\ \hline
                        V2 &    58.69   &   73.45       \\ \hline
                        V3 &    58.10   &   {\bf 71.64}       \\ \hline
                        V4 &    {\bf 57.88}   &   73.72       \\ \hline
    \end{tabular}
    \vspace{-5pt}
\end{table}
\subsection{ASR Experiments}
\par The acoustic model (AM) is trained using the Kaldi ``chain'' recipe \cite{Povey_ASRU2011} with LF-MMI \cite{Povey2016} as the minimization objective. 
A Bi-LSTM architecture\footnote{{KALDI-ROOT/egs/swbd/s5c/local/chain/tuning/run\_blstm\_6k.sh}} is used for the AM for Switchboard dataset, and a TDNN based architecture \footnote{KALDI-ROOT/egs/ami/s5b/local/chain/tuning/run\_tdnn\_1j.sh} is used for the AMI dataset. We consider experimentation using the individual head microphone (IHM) subset of the AMI dataset. A trigram language model, built on the training transcripts is used to perform first pass decoding to generate decoder lattices. The neural LMs are then used to rescore the $N$-best hypothesis list obtained from these lattices. The utterances are processed sequentially in CRNNLM and SessionLM~\cite{xiong2018session} rescoring. To rescore a given utterance hypotheses, a set of past $c$ decoded utterances are given as context input in CRNNLM.
\begin{table}[t]
    \centering
    \caption{Effect of (a) number of alternate hypotheses ($N$), and (b) length of context $c$, on the ASR performance (WER \%).}
    \vspace{-5pt}
    \label{tab:effect_nbest_context}    
    \begin{minipage}[b]{0.48\linewidth}
    \centering
    \centerline{ (a) WER vs $N$ ($c=2$)}
    \begin{tabular}{|c|c|c|}
        \hline
        $~N~$ & SWB & AMI  \\\hline
        20 & 13.5 & 19.0\\
        30 & 13.5 & 19.0\\
        50 & 13.3 & 18.8\\
        70 & 13.3 & 18.7\\
        100 & {\bf 13.2} & {\bf 18.6}\\\hline
    \end{tabular}    
    \end{minipage}
    \begin{minipage}[b]{0.48\linewidth}
    \centering
    \centerline{ (b) WER vs $c$ ($N=100$)}
     \begin{tabular}{|c|c|c|}
         \hline
         $~c~$ & SWB & AMI  \\\hline
         0 & 13.5 & 19.3\\
         1 & 13.2 & 18.7\\
         2 & 13.2 & 18.6\\
         3 & {\bf 13.1} & 18.6\\
         4 & {\bf 13.1} & {\bf 18.5}\\
         \hline
     \end{tabular}
    \end{minipage}    
    \vspace{-5pt}
\end{table}
\par First, we study the effect of the number $N$ of alternate hypotheses considered for rescoring. Table \ref{tab:effect_nbest_context}(a) shows the WER as a function of $N$. The WER is found to decrease with increase in $N$. 
The effect of the number $c$ of context utterances on WER is shown in table \ref{tab:effect_nbest_context}(b). The CRNNLM is trained with $c=2$ and it is varied during evaluation. For $c=0$ (no context), we consider the symbol `$<unk>$' as the context input. Compared with $c=0$, we see that the WER is better for $c>1$, indicating the usefulness of context. We see that, a context of $3$ utterances is sufficient and longer context does not necessarily result in WER improvement. This may be attributed to the local nature of the context embedding, which promotes repeated word sequences and continuity across utterances. Hence, we consider $N=100$ and $c=3$ for the following ASR experiments.
\begin{table}[t]
\centering
\caption{ASR performance for different LMs and CRNNLM variants, context $c=3$ and $N=100$.}
\vspace{-5pt}
\label{tab:wer_arch}
\begin{tabular}{|c|c|c|}
\hline
\multirow{2}{*}{LM} & \multicolumn{2}{c|}{WER \%} \\ \cline{2-3} 
                     & ~~~SWB~~~ & AMI        \\ \hline
                    \multicolumn{3}{|c|}{First pass} \\\hline  
                    trigram & 16.0 & 20.6 \\\hline
                    \multicolumn{3}{|c|}{Second pass (trigram + )} \\\hline  
                    Kaldi-RNNLM & 13.8 & 19.2 \\\hline
                    SessionLM~\cite{xiong2018session} & 13.4 & 19.0 \\\hline
                    V1 &    13.4           & 18.6        \\ \hline
                    V2 &    13.1           & 18.7         \\ \hline
                    V3 &    13.1           & 18.6         \\ \hline
                    V4 &    13.1           & 18.6        \\ \hline
                    \multicolumn{3}{|c|}{Third pass (trigram + Kaldi-RNNLM + )} \\\hline  
                    V3 & {\bf 12.6} & {\bf 18.2}\\\hline         
\end{tabular}
\vspace{-10pt}
\end{table}
\par The WER obtained using the different CRNNLM variants is shown in Table \ref{tab:wer_arch}. Rescoring using Kaldi-RNNLM is found to give an absolute improvement of upto $2.2\%$ compared to trigram LM, and using SessionLM gives upto $2.6\%$ in WER. The CRNNLMs are found to result in better WER than other models for both the datasets. Comparing V2-V4 with V1, we see that, relevance weighting of the context embedding helps in improving the WER. The architectures V2-V4 have similar performance for $N=100$, but for smaller $N$ we observed V3 to have slightly better performance. Table \ref{tab:wer_arch} also shows that CRNNLM rescoring of the N-best list generated using Kaldi-RNNLM (last row) improves the WER by absolute $1-1.2\%$ compared to Kaldi-RNNLM. Statistical significance analysis using bootstrap-CI approach \cite{1326009}, computed using the Kaldi tool {\it compute-wer-bootci}, showed the probability of improvement (POI) of the CRNNLM variants and the SessionLM to be $1$ compared to the Kaldi-RNNLM output. POI of the four CRNNLM variants V1-V4 compared to SessionLM is found to be $0.05, 0.90, 0.84, 0.92$ respectively, indicating that the output of CRNNLM architectures V2-V4 are statistically different from Kaldi-RNNLM and SessionLM significantly, and architecture V1 output is closer to SessionLM.
\section{Summary}\label{sec:concl}
In this paper, the use of conversation context from past utterances is shown to improve the LM and ASR transcription performance. The cross attention like architecture is found to extract local context features and also improve the next word prediction near the start of the sentence. The relevance weighting of the context features is also important and useful in improving the LM performance. An absolute improvement of $1-1.2\%$ in WER is obtained on two different types of conversations, telephone and multi-party meetings.
\section{Acknowledgments}
The authors would like to acknowledge the technical discussions with Abhishek Anand and Dr. Michael Free of BT Research that helped in shaping the paper.
\bibliographystyle{IEEEtran}
\bibliography{references}
\end{document}